\documentclass[twocolumn,prb,aps,showpacs,amsmath,amssymb,floatfix]{revtex4-1}

\usepackage{amsmath,amssymb,bm,epsfig}
\usepackage{dcolumn}
\usepackage{bm}
\usepackage{feynmf}

\begin{document}

\title{Reply to ``Comment on `Theory of Phonon-Assisted Adsorption in Graphene: Many-Body Infrared Dynamics' "}

\author{Sanghita Sengupta}

\affiliation{Institut Quantique and D$\acute{e}$partement de Physique, Universit$\acute{e}$ de Sherbrooke, Sherbrooke, Qu$\acute{e}$bec, Canada J1K 2R1}

\date{\today}
\begin{abstract}
  Based on a new self-energy for atom-phonon interaction, preceding Comment argues about the insufficiency of the mathematical techniques within the Independent Boson Model (IBM) to study physisorption in graphene membranes. In this Reply, we show that the new self-energy reported in the Comment is a perturbative expansion approximated for a 2-phonon process, severely divergent for membrane sizes larger than 100 nm and within its current mathematical form, ill-suited for investigating the physics of physisorption in graphene micromembranes. Additionally, we provide with further evidence of the adsorption rate within the IBM that reinforces the physical soundness of the mathematical techniques reported in Phys. Rev. B 100, 075429 (2019).
\end{abstract}
\maketitle
The main point of our paper \cite{sengupta2019} is: adsorption rate of low-energy atoms impinging normally on suspended, $\mu$m sized graphene membranes is \textit{finite}, approximately equal to the adsorption rate predicted by Fermi's golden rule. To arrive at this conclusion, we have used the Independent Boson Model (IBM) that captures the interaction  between the incoming atom and the phonons of the graphene membrane. Our mathematical technique for the calculation of the adsorption rate includes a self-energy formalism within the context of the IBM \cite{sengupta2019}.

In the Comment \cite{comment2019}, author questions the validity of our $\textit{finite}$ adsorption result and in that attempt, provides with a new self-energy for the atom-phonon interaction which includes additional terms that go beyond the IBM self-energy and are thus absent in our work \cite{sengupta2019}. Author then adapts our method for the calculation of the adsorption rate and extends it to this new self-energy. Within our formalism, he finds that the new self-energy fails to provide with a self-consistent solution. Author thus concludes that the failure of the new self-energy to give self-consistent solution must imply the invalidity of our mathematical formalism.

Additionally, while the Comment 
dismisses our method as invalid, it \textit{does not} provide with a mathematical technique that calculates the adsorption rate within this new self-energy. Thus, the Comment eludes the main point of our paper and remains inconclusive about the adsorption rate of incoming atoms.

In this Reply, we will first discuss some of the fundamentally important features of the new self-energy reported in the Comment. We will then show as to which one is invalid: our mathematical method to compute the adsorption rate or the new self-energy reported in the Comment. Finally, we will conclude our Reply with further evidence of the adsorption rate within the IBM that reinforces the physical soundness of the mathematical technique reported in Ref.~[\onlinecite{sengupta2019}].

Let us begin with \textit{our analysis} of the new self-energy reported in the Comment \cite{comment2019}. Throughout our Reply, we will refer to this self-energy as $\Sigma^{c}$. Eq.~(5) and Eq.~(6) of the Comment (see Ref.~[\onlinecite{comment2019}]) gives the new self-energy as\cite{comment2019}
\begin{widetext}
\begin{equation}\label{eq:secomment}
\begin{split}
 \Sigma^{c}(E) &=  g_{kb}^{2}\bigg[\sum_{q}\bigg(2\Lambda\lambda_{q}-2n_{q}^{2}\lambda_{q}^{2}\bigg)G^{\rm{IBM}}(E) - \sum_{p,q}\lambda_{p}\lambda_{q}\bigg(1+ 2n_{q}n_{p} + n_{q} +n_{p}\bigg)G^{\rm{IBM}}(E)\bigg] \\
 &\quad + g_{kb}^{2}\sum_{q}\bigg[\bigg\{n_{q}(2\Lambda\lambda_{q}+1)+2n_{q}^{2}\lambda_{q}^{2}\bigg\}G^{\rm{IBM}}(E+\omega_{q}) +\bigg\{(n_{q}+1)(1-2\Lambda\lambda_{q}) + 2n_{q}^{2}\lambda_{q}^{2}\bigg\}G^{\rm{IBM}}(E-\omega_{q})\bigg]\\
 &\quad +g_{kb}^{2}\sum_{q}\bigg[n_{q}\lambda_{q}^{2}(1-n_{q})G^{\rm{IBM}}(E+\omega_{q}+\omega_{q}) - (n_{q}+1)\lambda_{q}^{2}n_{q}G^{\rm{IBM}}(E-\omega_{q}-\omega_{q})\bigg]\\
 &\quad +g_{kb}^{2}\sum_{p,q}\bigg[n_{q}n_{p}\lambda_{q}\lambda_{p}G^{\rm{IBM}}(E+\omega_{q}+\omega_{p}) + (n_{q}+1)(n_{p}+1)\lambda_{p}\lambda_{q}G^{\rm{IBM}}(E-\omega_{q}-\omega_{p})\\
 &\quad - (n_{q}+1)n_{p}\lambda_{q}\lambda_{p}G^{\rm{IBM}}(E-\omega_{q}+\omega_{p}) - (n_{p}+1)n_{q}\lambda_{q}\lambda_{p}G^{\rm{IBM}}(E+\omega_{q}-\omega_{p})\bigg],\\
\end{split}
\end{equation}
\end{widetext}
where $g_{kb}$ is the vertex of atom-phonon interaction for a transition of atom from continuum to the bound state and $g_{bb}$ is the vertex of atom-phonon coupling for interaction in the bound state. $\lambda_{p} = g_{bb}/\omega_{p}$, $\Lambda =\sum_{p}\lambda_{p}$ and $n_{q}$ is the equilibrium phonon occupation number with Bose-Einstein distribution written as $n_{q} = 1/(e^{\omega_{q}/T}-1)$, where $\omega_{q}$ is the energy of the phonon with wave vector $q$ and $T$ is the temperature of the membrane \cite{comment2019} . $G^{\rm{IBM}}$ is the bound state Green's function written within the IBM (given by Eq.~(28) and Eq.~(29) in Ref.~[\onlinecite{sengupta2019}]) and $E= E_{k}+E_{b}$ with $-E_{b}$ as the bound state energy and $E_{k}$ is the incoming energy of the atom \cite{sengupta2019}.

$\Sigma^{c}(E)$ has additional terms compared to $\Sigma^{\rm{IBM}}$ as a result of the inclusion of non-commutativity of the phonon operator and displacement operator \cite{comment2019}. This non-commutativity was not addressed in our work \cite{sengupta2019}. Let us analyze $\Sigma^{c}$ with a special focus on the effects of the terms appearing as a result of the non-commutativity.  Below we provide with our points of disagreements concerning the form of $\Sigma^{c}(E)$ (given by Eq.~(\ref{eq:secomment})) and compare the same with the $\Sigma^{\rm{IBM}}$ reported in Ref.~[\onlinecite{sengupta2019}]. \\
(1.)\ While the Comment reports $\Sigma^{c}(E)$ as an $\textit{exact}$ closed-form expression for the atom self-energy to quadratic
order in the atom-phonon coupling O($g_{kb}^{2}$), we see that it is $\textit {in fact}$ a perturbative expansion in $\Sigma^{c} (E)$, truncated till a 2-phonon process. The definition of exact self-energy corresponds to summation of infinite number of Feynman diagrams which essentially implies the inclusion of the contribution from infinitely many phonons \cite{mahan}. By energy conservation, one can readily see that the propagators in Eq.~(\ref{eq:secomment}): $G^{\rm{IBM}}(E-\omega_{q}$) corresponds to 1-phonon emission with energy $\omega_{q}$ and $G^{\rm{IBM}}(E-\omega_{q}-\omega_{q})$, $G^{\rm{IBM}}(E-\omega_{q}-\omega_{p})$, $G^{\rm{IBM}}(E-\omega_{q}+\omega_{q})$ correspond to 2-phonon emission processes with energies $\omega_{q}$ and $\omega_{p}$. Thus $\Sigma^{c}$ is an $\textit{approximation}$ where the non-commutativity of the phonon and displacement operators has been incorporated upto 2-phonon processes. However, the Comment does not report $\Sigma^{c}$ as an approximation and furthermore, does not provide with the justification of such an approximation. $\Sigma^{\rm{IBM}}$, on the other hand, is also an approximation. It represents a 1-phonon self-energy which uses an $\textit{exact}$ propagator for the bound state $G^{\rm{IBM}}$ such that it includes all orders in the vertex $g_{bb}$, but is truncated till the first order phonon process in $g_{kb}$. This 1-phonon approximation in $g_{kb}$ has been justified for our model in Ref.~[\onlinecite{sengupta2019,sengupta2016}] within the context of relative magnitudes of the vertices $g_{kb}$ and $g_{bb}$ as $g_{kb}\ll g_{bb}$.

(2.) In the first line of Eq.~(\ref{eq:secomment}), we see that the bound state propagator is written as $G^{\rm{IBM}}(E)$. By energy conservation, this implies that the energy of the phonon is set to $\omega_{q} =0$. However, the vertex of interaction is written as $g_{kb}^{2}$. We remind ourselves that the definition of the vertex $g_{kb}$ for the model Hamiltonian in Refs.~[\onlinecite{sengupta2019},\onlinecite{clougherty2017T},\onlinecite{clougherty2014}] refers to the transition matrix element \cite{zhang2012,clougherty2014}

\begin{equation}\label{coupling}
g_{kb} = -\langle b,1_{q}|H_{i}|k,0\rangle,
\end{equation}
where $H_{i}$ is the Hamiltonian for atom-phonon interaction \cite{sengupta2019}. $|k,0\rangle$ represents the initial state of the atom $|k\rangle$ with energy $E_{k}$ and $|0\rangle$ is the graphene membrane in its ground state with no excitation. $|b\rangle$ is the final bound state with energy $-E_{b}$ and $|1_{q}\rangle$ represents excitation of 1 phonon with energy $\omega$ and wave vector $q$. Physically, Eq.~(\ref{coupling}) corresponds to the transition of atom from $|k\rangle$ to $|b\rangle$ via the emission of 1 phonon of energy $\omega$ and wave vector $q$ \cite{clougherty2014,zhang2012}. A similar definition exists for the vertex $g_{bb}$ which is the transition of the atom within bound states $|b\rangle$ via the emission of phonon of energy $\omega_{q}$ \cite{zhang2012,clougherty2014}. In other words, if the vertex $g_{kb}$ is used, it would imply an emission of phonon of energy $\omega_{q}$, which then appears via energy conservation in the expression for the bound state propagator $G^{\rm{IBM}}$. Therefore, the first line which is written with a propagator $G^{\rm{IBM}}(E)$, represents a process that involves $\textit{no}$ emission of phonon $\omega_{q}$, thus the use of the vertex $g_{kb}$ in such a situation is unjustified. Thus, within the definition of the vertices of atom-phonon coupling, the terms appearing from the non-commutativity of the phonon and displacement operators in the first line of Eq.~(\ref{eq:secomment}) are inaccurate.

\label{sec:se}

\begin{figure}[h]
\begin{center}
\includegraphics[width=\columnwidth]{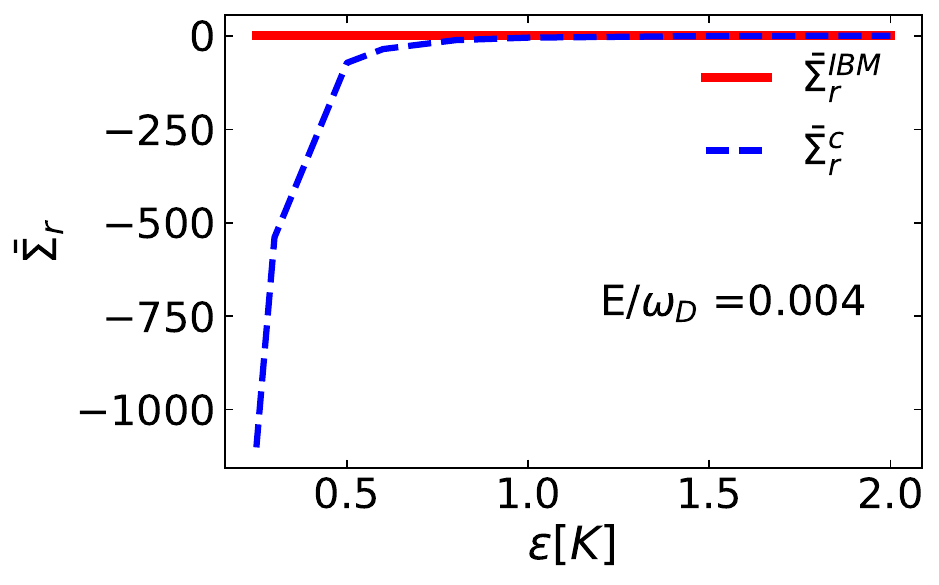} 
\end{center}
\caption{Variation of the real part of the dimensionless self-energy reported in Comment ($\bar{\Sigma}_{r}^{c}$) vs infrared cut-off $\epsilon$ (blue dashed line) for low-energy atom $E/\omega_{D} =0.004$. For $\epsilon<0.5$ K (membrane size $>$ 100 nm), the self-energy reported in the Comment starts to diverge severely with decreasing $\epsilon$ (increasing membrane sizes). Contrary to this, the real part of the self-energy within IBM\cite{sengupta2019} (red line) is well-behaved for the same ranges of IR cut-off (membrane sizes) and is around $\bar{\Sigma}_{r}^{\rm{IBM}}\sim 0.7$.} 
\label{fig:comparisonrelasigma}
\end{figure}

Armed with arguments (1.) and (2.), let us now proceed to understand the variation of the real part of the self-energy with the infrared (IR) cut-off $\epsilon$. The IR cut-off is related to the size of the graphene membranes by the relation $\epsilon = \hbar v_{s}/L$, where $v_{s}$ is the velocity of sound in graphene and $L$ is the size of the membrane. Thus a decreasing $\epsilon$ physically corresponds to increasing membrane sizes.  In what follows next, we will stick to the notations, labels and units consistent with the Comment \cite{comment2019}, unless otherwise mentioned.

In Fig.~\ref{fig:comparisonrelasigma}, we plot the variation of $\bar{\Sigma}^{c}_{r}$ (dimensionless real part of Eq.~(\ref{eq:secomment})) with $\epsilon$. For comparative purposes, we also show the variation of the dimensionless real part of the self-energy within IBM (real part of Eq.~(6) in Ref.~[\onlinecite{sengupta2019}]). Let us summarize our understanding and give further points of disagreements with the Comment as the following:

(3.) We note for $\epsilon \le 0.5$ K, $\bar{\Sigma}_{r}^{c}$ (blue dashed line) starts to diverge with decreasing $\epsilon$ (increasing size of graphene membranes). In comparison, $\bar{\Sigma}_{r}^{\rm{IBM}}$ within IBM (red line) is well-behaved for the same range of IR cut-off (size of membrane). Comment has reported these severe effects of IR divergence as mere $\textit{downward shifts}$ in the real part of the self-energy (not to mention, the absolute absence of physical justification for the presence of these IR divergences in a model of weak atom-phonon coupling). Mathematically, such a severe IR divergence signals the breakdown of the perturbation series, which implies that the perturbative expansion given by Eq.~(\ref{eq:secomment}) is ill-behaved for $\epsilon\leq 0.5$ K (membrane size $>$ 100 nm). Furthermore, Kinoshita-Lee-Nauenberg theorem \cite{kinoshita,lee,Bloch:1937pw} tells us that these IR divergences are physically unreal, hence proper resummations (non-perturbative techniques) need to be implemented to tackle these IR divergences with an effort to gain meaningful physical results. Our formalism of self-energy within the IBM in Ref.~[\onlinecite{sengupta2019}] is $\textit{in fact}$, a resummation technique that was implemented to tackle the severe IR divergences which appear with the inclusion of the effects from the atom-phonon coupling in the bound state \cite{sengupta2019,sengupta2016}. 

(4.) Previously in point (2.), we mentioned about the inaccuracy of the terms appearing as a result of non-commutativity of phonon and displacement operators in the first line in  Eq.~(\ref{eq:secomment}). It is to be noted that there is a contribution to the leading order divergence in $\bar{\Sigma}_{r}^{c}$ in the limit of $\epsilon\rightarrow 0$ that originates from the term
\begin{equation}\label{infinite}
-\sum_{q}2n_{q}^{2}\lambda_{q}^{2}G^{\rm{IBM}}(E) = -\frac{1}{\epsilon^{3}}\bigg[\frac{2g_{bb}^{2}T^{2}}{3}G^{\rm{IBM}}(E)\bigg] \rightarrow -\infty.
\end{equation}


With the knowledge of points (3.) and (4.), let us now state our final points of disagreements with the arguments provided in the Comment for the invalidity of our method for the calculation of the adsorption rate. 

(5.) Utilizing the real ($\Sigma_{r}$) and imaginary part of the self-energy ($\Sigma_{i}$), the adsorption rate $\Gamma$ within our method is given as
\begin{equation}\label{ar}
\Gamma \approx -2\mathcal{Z}\Sigma_{i}(E_{p}),
\end{equation}
where the quasiparticle weight $\mathcal{Z}$ is
\begin{equation}\label{Z}
\mathcal{Z} = \bigg(1-\frac{\partial\Sigma_{r}(E)}{\partial E}\bigg|_{E=E_{p}}\bigg)^{-1},
\end{equation}
$E_{p}$ is the quasiparticle energy that can be solved via
\begin{equation}\label{energy}
E_{p}-E_{k} = \Sigma_{r}(E_{p}).
\end{equation}
Using the real part of the self-energy $\Sigma_{r}^{c}$ (given by real part of Eq.~(\ref{eq:secomment})), Comment attempts to find a graphical solution to Eq.~(\ref{energy}). For low-energy atoms, author finds no self-consistent solution in the range of $\epsilon\le 0.4$ K (see Fig.~6 in Comment\cite{comment2019}). This failure is \textit{not because} of the change in sign of the curvature of the real part of the self-energy (as reported by the Comment\cite{comment2019}), but rather from an infinite (divergent) self-energy plugged into the rhs of Eq.~(\ref{energy}) (see the IR divergent behavior of $\Sigma_{r}^{c}$ in Fig.~\ref{fig:comparisonrelasigma}). In contrast, the IBM self-energy which is well-behaved for similar ranges of IR cut-off, succeeds to give self-consistent solution to Eq.~(\ref{energy}), also evident from the Fig.~5 and Fig.~6 of the Comment\cite{comment2019}. 
\begin{figure}[h]
\begin{center}
\includegraphics[width=0.9\columnwidth]{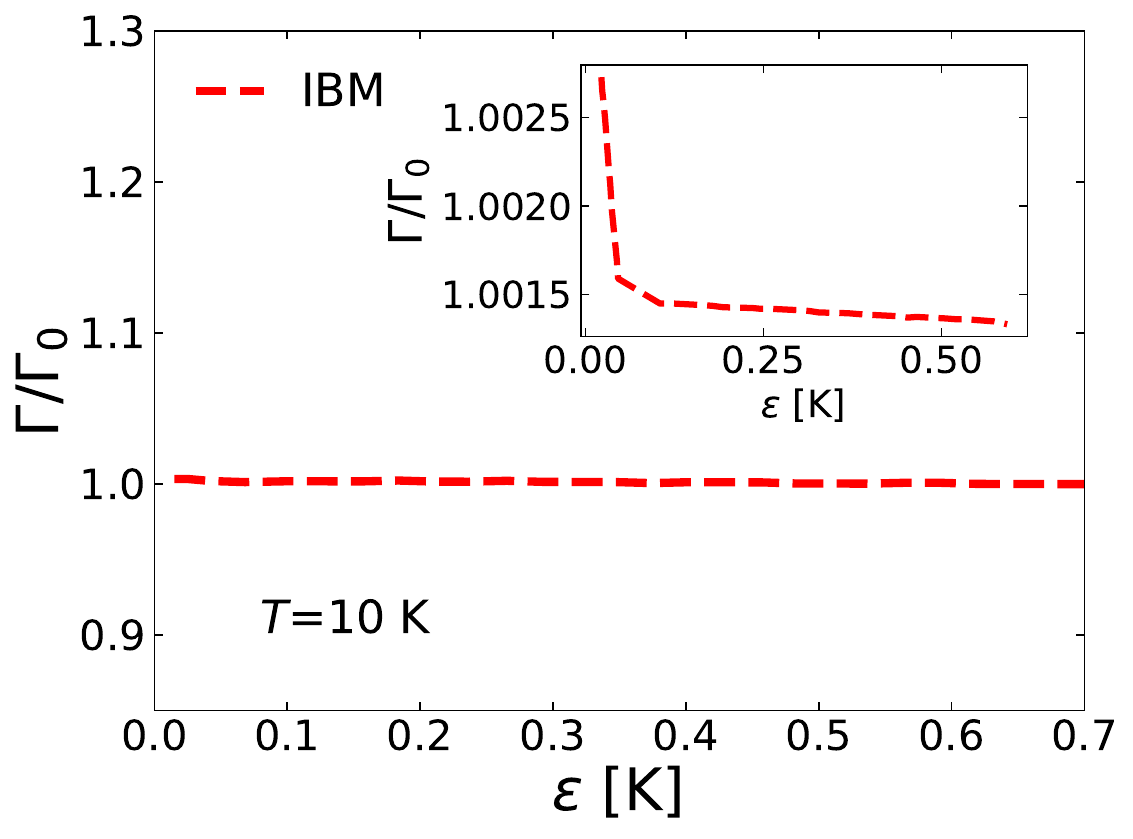} 
\end{center}
\caption{Within the self-energy in IBM \cite{sengupta2019}, we plot a variation of the normalized adsorption rate $\Gamma/\Gamma_{0}$ vs IR cut-off $\epsilon$, $\Gamma_{0}$ is the Fermi's Golden rule result. Little to no variation is seen with $\epsilon$ (i.e with membrane sizes of range 100 nm $\sim$ 10 $\mu$m). Inset shows the variation for very low $\epsilon$. Small enhancement (0.27 $\%$ of $\Gamma_{0}$) in adsorption rate is seen with decreasing $\epsilon$ (increasing membrane sizes), which is related to the enhancement of probability of emission of low-energy thermal phonons (obeying Bose-Einsten distribution) at finite temperatures.} 
\label{fig:ibmfullrate}
\end{figure}

Before we conclude, let us provide with an additional calculation of the adsorption rate of low-energy atoms using the self-energy within IBM. This calculation clarifies some of the inaccurate representation of our results provided in the Comment. In Fig.~\ref{fig:ibmfullrate}, we have shown the variation of the normalized adsorption rate $\Gamma/\Gamma_{0}$ as a function of the IR cut-off ($\epsilon$). Here, $\Gamma_{0}$ is the Fermi's golden rule result. One can see that for a large range of IR cut-offs that corresponds to membrane sizes 100 nm $\sim$ 10$\mu$m, the adsorption rate $\Gamma\approx\Gamma_{0}$. In the very low IR cut-off regime ($\epsilon\le 0.1~$K), we find small increments to the adsorption rate ($\Gamma$ is still within 0.27 $\%$ of $\Gamma_{0}$, see inset of Fig.~\ref{fig:ibmfullrate}). Comment mentions there is a divergence in our results which becomes apparent for $\epsilon\leq 0.1$ K. We point out that this increment in the adsorption rate physically represents the temperature effect of the Bose-Einstein distribution obeyed by the thermal phonons in the graphene membrane. As temperature (or size of the membrane) is increased (decreasing $\epsilon$), there is an enhancement of the probability of emission of low-energy thermal phonons, leading to an increase in adsorption rate\cite{sengupta2016}.

In conclusion, Comment\cite{comment2019} has reported a new self-energy which is a perturbative expansion that includes the non-commutativity of the phonon operator and displacement operator, approximated till a 2-phonon process. This $\textit{approximation}$ (although reported as an exact method in the Comment\cite{comment2019}) is ill-behaved for low IR cut-offs (large membrane sizes) and suffers from severe IR divergences; tracing back to the original IR problem of the model where perturbative treatment of the self-energy generally leads to IR divergent self-energy, signalling the need for resummation to be performed on the perturbative series expansion. Our method within the IBM \cite{sengupta2019} is a resummation technique that was indeed formulated as a measure to tackle these IR divergences that arise in the perturbative treatment of the problem. Quite naturally, the IR structure of the $\Sigma^{\rm{IBM}}$ and $\Sigma^{(c)}$ are starkly different with the merit of $\Sigma^{\rm{IBM}}$ being well-behaved for micromembranes of graphene samples. An IR safe self-energy is a general as well as crucial requirement for the calculation of the adsorption rate. As the self-energy reported in the Comment \cite{comment2019} is severely IR divergent for IR cut-off less than 0.5K (corresponding to membrane sizes larger than 100 nm), naturally it $\textit{fails}$ to predict adsorption rates for graphene membrane sizes larger than 100 nm. Additionally, the Comment \cite{comment2019} has also misplaced the atom-phonon vertex on one of the non-commutativity terms in $\Sigma^{c}$ which has resulted in a leading order IR divergence; these terms are however, not allowed in our model \cite{sengupta2019,zhang2012,clougherty2014} within the definition of the atom-phonon vertices. In the absence of such terms, the perturbative series expansion in the Comment\cite{comment2019} is still weakly IR divergent, urging the need of a resummation to be performed on the perturbative series. Unless proper resummations are performed with appropriate placement of atom-phonon coupling, this divergent self-energy reported in the Comment \cite{comment2019} remains unsuitable for physisorption studies in graphene micromembranes. In contrast, the self-energy within IBM, reported in Ref.~[\onlinecite{sengupta2019}] is a resummed self-energy \cite{sengupta2019,sengupta2016}, well-behaved for the same range of IR cut-off (see Fig.~\ref{fig:comparisonrelasigma}) and is conclusive about the adsorption rate for a large range of membrane sizes 100 nm $\sim$ 10$\mu$m, suitably capturing the physics of temperature and finite size effects. Finally, let us address the concluding remark of the Comment\cite{comment2019} which states that the difference between the finite adsorption rate predicted in Ref.
[\onlinecite{sengupta2019}] with the zero adsorption rates predicted in Ref.~[\onlinecite{clougherty2017T},\onlinecite{clougherty2014}] is due to the self-energy used within the IBM. We \textit{strongly disagree} with this remark. Within the simple model of IBM, our original work had shown that the zero adsorption rate is only possible if one considers (i) contribution to adsorption rate from the long time regime, where the effects of Franck-Condon factor sets in and (ii) neglects the effects of thermal phonon emission. Points (i) and (ii) are indeed the regime of study in Ref.~[\onlinecite{clougherty2014}] and Ref.~[\onlinecite{clougherty2017T}], respectively. However, if we consider the contribution to adsorption rate from full time regime and do not neglect the effects of thermal phonon emission (which is imminent for finite temperature physics), the adsorption rate will be $\textit{finite}$, equal to Fermi's Golden rule \cite{sengupta2019, LJ2011,LJreply}, validating the IR-divergence cancellation predicted by the Bloch-Nordsieck theorem \cite{Bloch:1937pw}. It would be interesting to know if $\Sigma^{c}$ reported in the Comment, would also give the same finite adsorption rate as $\Sigma^{\rm{IBM}}$ when the following improvements are made within $\Sigma^{c}$, namely: (i) the atom-phonon vertices in the non-commutativity terms are not misplaced, (ii) a full resummation (non-perturbative) formalism is performed on the weakly IR divergent perturbative series, by including the contribution of infinitely many low-energy phonons and not just 2-phonons and (iii) contribution to adsorption rate is inclusive for a full-time regime without neglecting thermal phonon emissions.

\section*{acknowledgments}
I am grateful to Professor Ion Garate for an insightful discussion and his valuable advice. This work was funded by the Canada First Research Excellence Fund. 

\bibliographystyle{apsrev4-1}
\bibliography{reply}
\end{document}